\documentclass[12pt,aps,prl,floatfix,tightenlines]{revtex4}  
\pdfoutput=1
\usepackage[pdftex]{hyperref}
\usepackage{graphicx}
\usepackage{color}
\usepackage{amsmath}
\usepackage{latexsym}

\begin{document}

\title{Chirped-pulse interferometry with finite frequency correlations}

\author{K.J. Resch}\email{kresch@iqc.ca}
\author{R. Kaltenbaek} 
\author{J. Lavoie}
\author{D.N. Biggerstaff}
\affiliation{Institute for Quantum Computing and Department of Physics \& 
Astronomy, 200 University Ave. W, Waterloo, N2L 3G1, Canada}

\begin{abstract}
Chirped-pulse interferometry is a new interferometric technique
encapsulating the advantages of the quantum Hong-Ou-Mandel
interferometer without the drawbacks of using entangled photons.
Both interferometers can exhibit even-order dispersion cancellation
which allows high resolution optical delay measurements even in thick
optical samples.  In the present work, we show that finite frequency
correlations in chirped-pulse interferometry and Hong-Ou-Mandel
interferometry limit the degree of dispersion cancellation. Our
results are important considerations in designing practical devices
based on these technologies.
\end{abstract}
\maketitle

\section{INTRODUCTION}

Interference is a fundamental characteristic shared by
classical and quantum theories of light, and is used to make
the most sensitive measurements of quantities such as distance
and time.  Although interferometry has long played an important
role in physics (eg. from Ref.~\cite{Michelson1887} to
Ref.~\cite{Abbott2005}~\hspace{-1mm}), recent work has raised
the question as to how interferometers harnessing quantum
effects can provide advantages over purely classical ones.
Experiments with entangled photons have demonstrated a wide
variety of interference effects that had not previously been
seen in classical devices. Prominent examples include automatic
dispersion
cancellation~\cite{Steinberg1992a,Steinberg1992b,Franson1992}~\hspace{-1mm},
phase-insensitive interference~\cite{Hong1987}~\hspace{-1mm},
nonlocal interference~\cite{Franson1989,Ou1990}~\hspace{-1mm},
ghost imaging~\cite{Pittman1995}~\hspace{-1mm} \& ghost
diffraction~\cite{Strekalov1995}~\hspace{-1mm}, phase
super-resolution~\cite{Lee2002,Giovannetti2004,Mitchell2004,Walther2004}~\hspace{-1mm},
and phase
super-sensitivity~\cite{Yurke1986,Rarity1990}~\hspace{-1mm}.

More recently, there has been an effort to observe analogous
effects in classical optical interferometers.  Ghost imaging
\cite{Bennink2002,Ferri2005}~\hspace{-1mm}, automatic
dispersion cancellation
\cite{Resch2007,Resch2007b,Kaltenbaek2008,Lavoie2009}~\hspace{-1mm},
phase super-resolution \cite{Resch2007a}~\hspace{-1mm}, and
phase-insensitive interference
\cite{Kaltenbaek2008,Kaltenbaek2009} have been observed in
optical systems exploiting purely classical correlations,
instead of entanglement. Classical systems have been devised
which can exhibit Bell-like correlations, albeit not fulfilling
the conditions of a Bell experiment and thus without
implications for local hidden variable models
\cite{KFLee2002,KFLee2004,KFLee2009}~\hspace{-1mm}.  Other
methods for dispersion cancellation in classical systems have
been described theoretically
\cite{Erkmen2006,Banaszek2007}~\hspace{-1mm}.

From a fundamental perspective this body of work aims to
distinguish those cases where quantum effects give a true
advantage over classical systems from those where they do not.
From a more applied point of view, quantum effects are
technically challenging to observe, often requiring generation
of fragile entangled states and sensitive detection. Signals
are often very weak requiring long integration times.
Compounding the last issue, many schemes rely on the
correlations in a specific entangled state; one cannot simply
increase the number of photons (say by running two entangled
states through the interferometer simultaneously) without
ruining the requisite quantum correlations.  In comparison,
many classical schemes offer large signals which are easy to
detect, and furthermore can often be increased simply by
turning up the power of a laser.

Material dispersion is a limiting issue in low-coherence, or
white-light, interferometry.  If dispersive materials are
placed in one arm of a two-path interferometer (such as a
Michelson interferometer), the resulting interferogram will
lose contrast and resolution. In essence, the different
frequency components of the white light disagree on when the
interferometer path lengths are balanced.  The quantum,
Hong-Ou-Mandel interferometer (see Fig.~1a) using
frequency-entangled photon pairs produces an interferogram that
exhibits rather surprising robustness against dispersive
broadening. This device relies on frequency-entangled photon
pairs.  One photon from the pair travels through a dispersive
medium, while the other passes through an adjustable delay.
After recombination on a 50/50 beamsplitter, the coincidences
are recorded as a function of the delay. The coincidence rate
shows a dramatic dip when the group delays of the two paths are
balanced to within the coherence time of the light. This can be
as short as a few fs \cite{Nasr2008}~\hspace{-1mm}.
Theoretically, the width of the coincidence dip has been shown
to be insensitive to all even-orders of dispersion
\cite{Steinberg1992b}~\hspace{-1mm}. Experiments have confirmed
that the dip is much less susceptible to broadening than
white-light interferometry \cite{Steinberg1992a}~\hspace{-1mm}.
It has been proposed that optical coherence tomography, a
medical imaging technique based on low-coherence
interferometry, could benefit from automatic dispersion
cancellation \cite{Abouraddy2002,Nasr2003}~\hspace{-1mm}.
However, the reliance on entangled photons presents a
significant barrier to rapid signal acquisition.

We have recently demonstrated an interferometric technique that
produces an interferogram with the same advantages as the HOM
interferometer with frequency-entangled photon pairs, yet
requires no entanglement at all
\cite{Kaltenbaek2008}~\hspace{-1mm}.  This technique relies on
pairs of classically frequency anticorrelated beams, created by
using pairs of oppositely chirped laser pulses.  We referred to
the device as chirped-pulse interferometry (CPI).  A schematic
for CPI is shown in Fig.~2a.  Pairs of oppositely chirped
pulses enter into different input ports of a cross-correlator.
As in the HOM, the cross-correlator contains a dispersive
element in one arm, and a delay in the other.  The intensity of
a narrow bandwidth of SFG created in a nonlinear crystal is
detected on a standard photodiode. This intensity as a function
of delay is the CPI interferogram.  In
Ref.~\cite{Kaltenbaek2008} all of the important features of HOM
interference, automatic dispersion cancellation, robustness
against loss, phase insensitivity, and enhanced resolution were
demonstrated. Furthermore, the signal level measured was
roughly seven orders of magnitude larger than that achievable
in the quantum device with state-of-the-art photon sources.  We
have since demonstrated the effectiveness of the device in
imaging a multi-interface sample
\cite{Lavoie2009}~\hspace{-1mm}.  We have shown that
modifications to the device can produce analogous
interferometry signatures to quantum beating, the HOM peak, and
2-photon phase super-resolution
\cite{Kaltenbaek2009}~\hspace{-1mm}.

Dispersion cancellation in these devices has been described
theoretically in the literature
\cite{Steinberg1992b,Abouraddy2002,Kaltenbaek2008}~\hspace{-1mm},
but these descriptions assume perfect frequency correlations.
Perfect correlations in the HOM interferometer would require
down-conversion pump lasers with infinitely narrow spectra,
while perfect correlations in CPI would require infinitely long
chirped pulses. Of course, these limits are unattainable in
practice and here we investigate, theoretically, the limits to
dispersion cancellation in the presence of imperfect
correlations.  The goal of this work is to develop rules of
thumb describing the reduction in dispersion sensitivity
afforded by a given level of correlation. These considerations
will be important in designing practical interferometers which
take advantage of dispersion cancellation.

\section{DISPERSION CANCELLATION IN A HONG-OU-MANDEL INTERFEROMETER WITH FINITE CORRELATIONS}

Automatic even-order dispersion cancellation occurs in the
two-photon Hong-Ou-Mandel interferometer when the photon pairs
are frequency entangled.  In the literature
\cite{Steinberg1992b,Abouraddy2002}~\hspace{-1mm}, the quantum
state produced by parametric down-conversion from a narrow band
pump laser has been described as,
\begin{eqnarray}
|\psi\rangle = \int d\Omega f(\Omega)
|\omega_0+\Omega\rangle_1|\omega_0-\Omega\rangle_2.
\end{eqnarray}
In this state, the sum of the frequencies of the pair of photons is
fixed at $2\omega_0$ while the individual photons may be
individually broadband, depending on the function $f(\Omega)$. It
has been shown that with such a state, the HOM interference dip is
completely insensitive to even-order dispersion. However, this state
is an approximation to what can actually be achieved.  Any real
laser has a finite bandwidth and passes on this frequency
uncertainty to the photon pairs.

To obtain some intuition regarding dispersion cancellation with
imperfect frequency anticorrelations, we describe the state of a
pair of photons with adjustable frequency anticorrelations using,
\begin{eqnarray}
|\psi\rangle = \int d\omega_1 \int d\omega_2 f(\omega_1,\omega_2)
|\omega_1\rangle_1|\omega_2\rangle_2,
\end{eqnarray}
with the function, $f(\omega_1,\omega_2)$,
\begin{eqnarray}
f(\omega_1,\omega_2)=e^{-\frac{(\omega_1-\omega_0)^2}{2\sigma^2}}
e^{-\frac{(\omega_2-\omega_0)^2}{2\sigma^2}}
e^{-\frac{(\omega_1+\omega_2-2\omega_0)^2}{2\sigma_c^2}},
\end{eqnarray}
where $\sigma$ is the bandwidth of the photons, and $\omega_0$ the centre frequency.
If the photons were created via a parametric down-conversion
process, the parameter $\sigma_c$ plays the role of the bandwidth of the
pump laser and controls the strength of the frequency correlations.
If $\sigma_c \gg \sigma$, then $f(\omega_1,\omega_2) \approx
f_1(\omega_1)f_2(\omega_2)$, i.e., the bandwidth function is separable and the photons have no frequency
correlations.  In the opposite limit, where $\sigma_c \rightarrow 0$
then $e^{-\frac{(\omega_1+\omega_2-2\omega_0)^2}{2\sigma_c^2}}$
$\rightarrow$ $\delta(\omega_1+\omega_2-2\omega_0)$ and the photons
have perfect frequency anti-correlations.

\begin{figure}
\label{HOM}
\begin{center}
\includegraphics[width=1 \columnwidth]{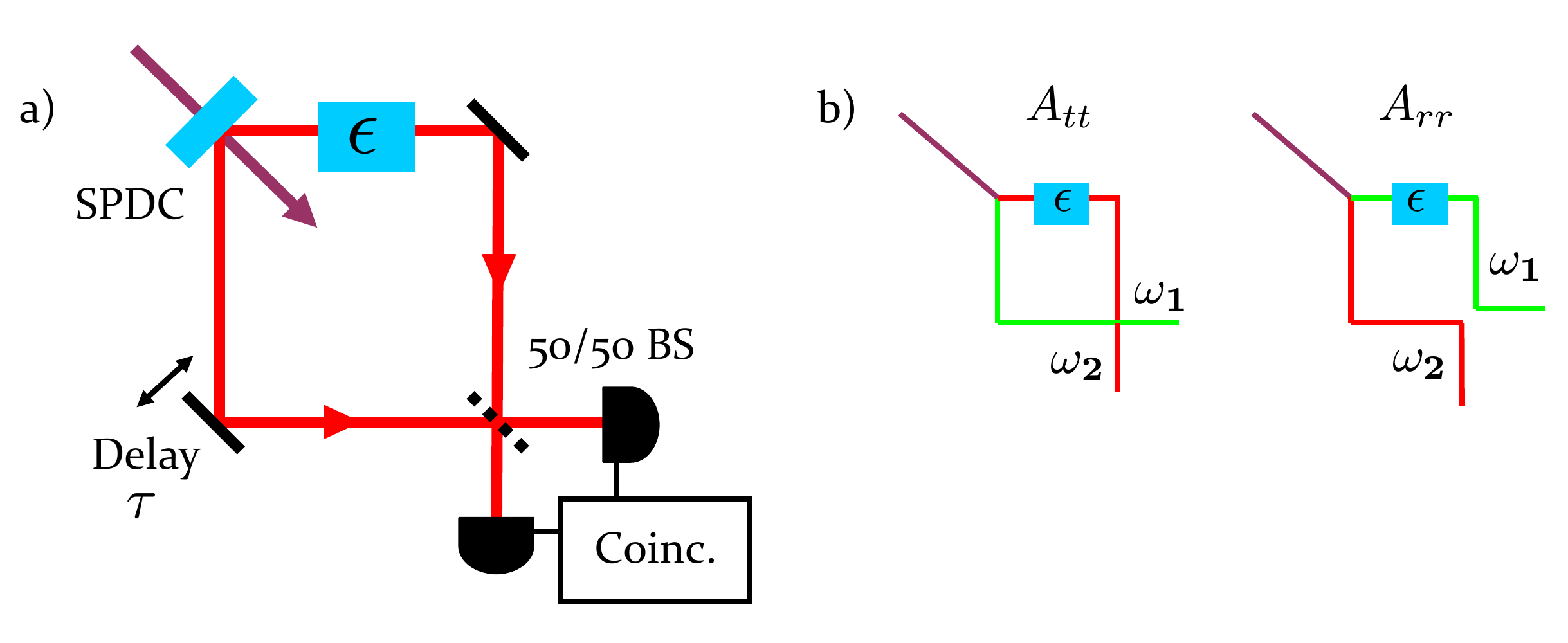}
\caption{Schematic of Hong-Ou-Mandel interferometry.  a) Photon
pairs, typically, though not necessarily, created via parametric
down-conversion are emitted into two different spatial modes.  One
mode passes through a delay arm and the other passes through a
sample.  The photon pairs are recombined coherently at a 50/50
beamsplitter after the delay and samples.  Photon counting detectors
are placed in the output modes of the beamsplitter.  The signal from
the interferometer is the coincidence rate between those detectors
as a function of the delay, $\tau$. b) There are two amplitudes
leading to the detection of photons with frequencies, $\omega_1$ and
$\omega_2$.  These amplitudes correspond to the Feynman paths shown, where
either both photons are reflected at the beamsplitter or both
photons are transmitted.  The HOM signal is built up by the coherent
addition of these amplitudes and the incoherent addition over all
possible frequency pairs.}
\end{center}
\end{figure}

The Hong-Ou-Mandel interferometer is shown in Fig.~1a where a
dispersive element is located in one arm.  The HOM interference
signal is given by the coincidence rate between the two slow
square law detectors and is given by,
\begin{eqnarray}
\label{coinc} C(\tau)\propto \int \!\!\! \int d\omega_1 d\omega_2
|A_{tt}(\omega_1, \omega_2,\tau)+A_{rr}(\omega_1,\omega_2,\tau)|^2,
\end{eqnarray}
where $A_{tt}$ ($A_{rr}$) is the amplitude where both photons
are transmitted (reflected) at the beamsplitter.  These are
depicted in Fig.~1b.  Here, we are only interested in the width
of the HOM interference dip so we ignore the path lengths of
the interferometer since they just lead to a time-delay offset.
We model the material as subjecting the light to a pure
quadratic phase shift $\phi(\omega)=\epsilon
(\omega-\omega_0)^2$, where $\epsilon$ is a constant
characterizing the strength of the dispersion. Note that we
have dropped the group delay term which again just leads to an
offset.  The amplitudes for the two paths are,
\begin{eqnarray}
A_{tt}= f(\omega_1,\omega_2)e^{i \epsilon(\omega_2-\omega_0)^2}
e^{i\omega_1\tau}\\
A_{rr}= -f(\omega_1,\omega_2)e^{i \epsilon(\omega_1-\omega_0)^2}
e^{i\omega_2\tau},
\end{eqnarray}
where we have used the fact that
$f(\omega_1,\omega_2)=f(\omega_2,\omega_1)$. Integrating
Eq.~\ref{coinc} with these amplitudes in yields
\begin{eqnarray}
\label{homdip}
C(\tau)\propto 1-\sqrt{\frac{2\sigma^2+\sigma_c^2}{2\sigma^2+\sigma_c^2+\epsilon^2 \sigma^4 \sigma_c^2}}e^{-\frac{\tau^2\sigma^2(2\sigma^2+\sigma_c^2)}{2(2\sigma^2+\sigma_c^2+\epsilon^2\sigma^4\sigma_c^2)}}.
\end{eqnarray}

From this expression, we see that the HOM signal is a Gaussian
function of the time delay, $\tau$.  The RMS width of the HOM
signal, $\tau_{HOM}$,
\begin{eqnarray}
\tau_{HOM}=\frac{1}{\sigma}\sqrt{1+\frac{\epsilon^2\sigma^4\sigma_c^2}{2\sigma^2+\sigma_c^2}}.
\end{eqnarray}
It is useful to look at this expression in a few limits.  If
the dispersion of the material is 0, i.e., $\epsilon=0$, then
$\tau_{HOM}=\frac{1}{\sigma}$.  In other words, the width of
the HOM dip is approximately the coherence time of the photon.
If the photons have perfect frequency correlations, i.e.,
$\sigma_c = 0$, then the width of the HOM dip is again given by
the coherence time. In other words, if the photons have
perfectly anticorrelated frequencies, then the width of the HOM
dip in the presence of pure second-order dispersion is exactly
the same as if there was no dispersion.  This agrees with the
conclusions of
\cite{Steinberg1992b,Abouraddy2002}~\hspace{-1mm}. Note that if
there are no frequency correlations, $\sigma_c \rightarrow
\infty$, then the dip is sensitive to the dispersion with
$\tau_{HOM}\rightarrow\frac{1}{\sigma}\sqrt{1+\epsilon^2\sigma^4}$.

Now we can consider the effect of imperfect correlations on the
dispersion cancellation.  Consider the limit where the photon
correlations are strong, but imperfect, $\sigma_c \ll \sigma$.  In
this case, the width is given by
\begin{eqnarray}
\label{approxHOMwidth}
\tau_{HOM}\approx\frac{1}{\sigma}\sqrt{1+\epsilon^2\sigma^4\frac{\sigma_c^2}{2\sigma^2}}.
\end{eqnarray}
Here it can be seen that the sensitivity to dispersion is
reduced by a factor, $\frac{\sigma_c^2}{2\sigma^2}$. Assuming
the photons are created by down-conversion, this reduction is
roughly $\frac{\tau_{photon}^2}{\tau_{pump}^2}$, where
$\tau_{photon}$ is the coherence time of a down-conversion
photon and $\tau_{pump}$ is the coherence time of the pump
laser.  In other words the dispersion, $\epsilon$, is reduced
to $\epsilon\frac{\tau_{photon}}{\tau_{pump}}$ as a result of
the frequency correlations. Note that even quantum dispersion
cancellation is not perfect when the correlations are finite.
However, in practice, it is straightforward to create
down-converted photon bandwidths that are millions of times
larger than the bandwidth of the pump laser
\cite{Kwiat1993,Nasr2008}~\hspace{-1mm}, thus the sensitivity
to second-order dispersion can be reduced to a negligible
level. One should keep in mind that automatic dispersion
cancellation applies only to even-order dispersion; once the
cubic (or third-order) dispersion becomes dominant, further
reductions in sensitivity to second-order dispersion will not
significantly aid signal quality.

\section{DISPERSION CANCELLATION IN CHIRPED-PULSE INTERFEROMETRY WITH FINITE CHIRP}

Chirped-pulse interferometry (CPI) is a new technique designed
to produce the metrological signature of the Hong-Ou-Mandel
interferometer without the need for entanglement or single
photons. A schematic of the chirped-pulse interferometer is
shown in Fig.~2a.  Two laser pulses are created with equal but
opposite chirps; we refer to them as chirped (where the blue
lags the red) and anti-chirped (where the red lags the blue).
They are injected into different ports of the cross-correlator
and recombined in a sum-frequency generation (SFG) after
traveling through two different paths. A narrow spectrum of the
SFG is detected on a square-law detector. The intensity of the
SFG as a function of delay is the CPI signal.

\begin{figure}
\label{CPI}
\begin{center}
\includegraphics[width=1 \columnwidth]{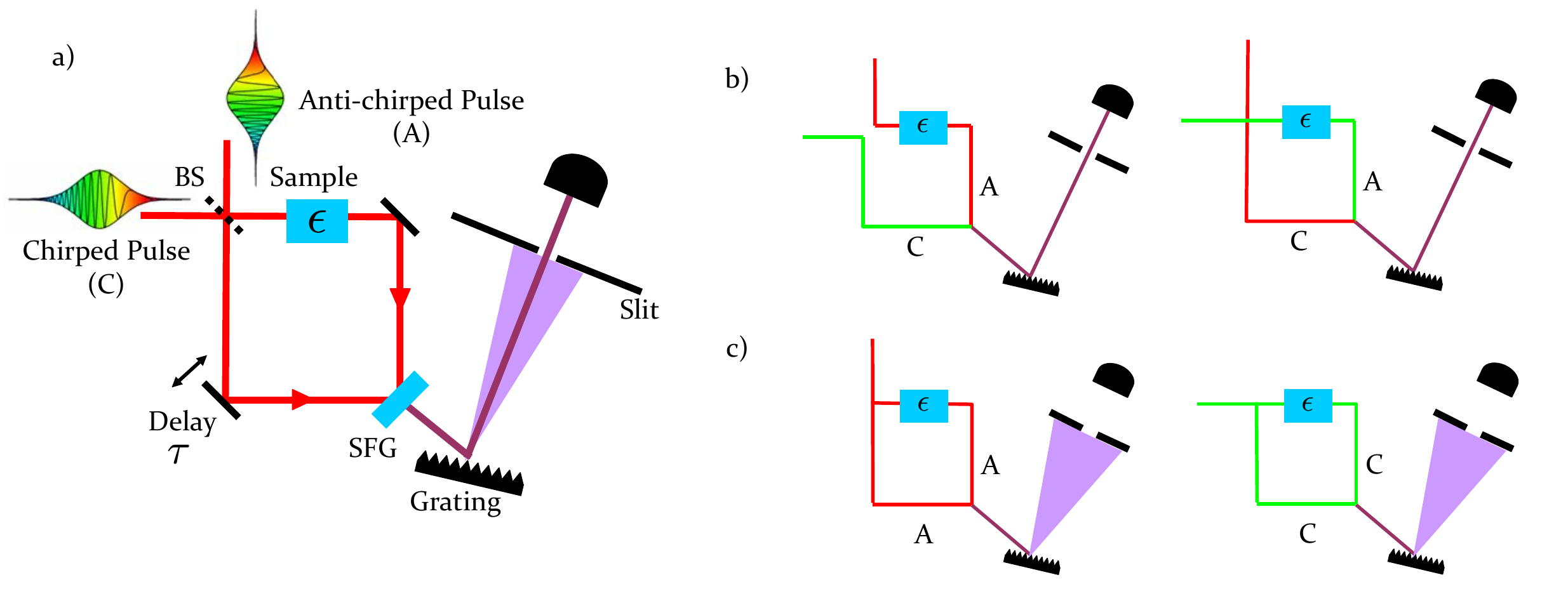}
\caption{Schematic of chirped-pulse interferometry.  a)
Chirped-pulse interferometer.  A pair of oppositely-chirped laser
pulses are combined coherently at a 50/50 beamsplitter.  A variable
delay arm is placed in one of the modes (a reference) while a sample
is placed in the other arm.  (In this paper, the sample will be
assumed to introduce a pure quadratic phase,
$\phi(\omega)=\epsilon(\omega-\omega_0)^2$).  After the sample,
the light in the reference and sample arms are used to create
sum-frequency generation (SFG) in a nonlinear material.
The narrow band of the SFG created by the cross-correlation of the
chirped and anti-chirped beams is detected with a
square-law detector. The detected intensity as a function of the
delay, $\tau$, is the CPI signal.  b) The two Feynman paths leading
to narrow-band SFG.  These involve contributions where the chirped
pulse passes through the sample arm and the anti-chirped pulse
passes through the reference or vice versa.  c) The two paths
leading to broad band SFG.  These contributions are from the
autocorrelations where the chirped (or antichirped) pulses alone
create SFG.  Contributions to the signal from these processes can be almost
entirely removed from spectral filtering when the pulses are stretched to many times their
transform-limited duration.  In the calculations presented here, we
ignore these processes.}
\end{center}
\end{figure}

The nonlinear detection in CPI relies on sum-frequency
generation, a second-order nonlinear process.  In such
processes, the nonlinear polarization, $P_{NL}(t)$ in the
medium depends on the product of the two driving fields:
\begin{equation}
\label{nonlinearpol}
P_{NL}(t) \propto \chi^{(2)} E_1(t) E_2(t)
\end{equation}
This nonlinear polarization can reradiate light at frequencies
corresponding to its frequency components.  We make the
assumption that the nonlinearity is fast so that $\chi^{(2)}$
can be treated as a constant.  We will also make the assumption
that the driving fields are not depleted and that perfect
phase-matching over all driving frequencies is achieved.  We
make the approximation that the radiated field is proportional
to the polarization and the slowly varying amplitude
approximation \cite{BoydNLO} and rewrite Eq.~\ref{nonlinearpol}
to obtain an expression for the SFG electric field amplitude in
the frequency domain,
\begin{equation}
E_3(\omega) \approx \int d\omega' E_1(\omega')E_2(\omega-\omega')
\end{equation}
The positive frequency electric field amplitudes for a linearly
chirped pulse can be written,
\begin{equation}
E(\omega;A)=E_0e^{\frac{(\omega-\omega_0)^2}{2\sigma^2}}e^{iA(\omega-\omega_0)^2},
\end{equation}
where $\sigma$ is the RMS bandwidth of the field and $A$
determines the strength and sign of the chirp.  CPI relies on
strong frequency correlations which come from chirping the
pulses to many times their transform-limited pulse duration,
i.e., $A \gg \frac{1}{\sigma^2}$.

We overlap the oppositely chirped pulses on the input beam
splitter of our interferometer. In one arm of the
interferometer the pulses experience a relative time delay
$\tau$ with respect to the other arm. The field in this delay
arm can be written as:
\begin{equation}
E_1(\omega,\tau) = \left[E(\omega;A)+E(\omega;-A)\right] e^{i \tau
\omega}.
\end{equation}
In the second arm of the interferometer the light passes
through a sample with purely quadratic dispersion (i.e., we
ignore the group delay which just leads to an offset of the
intereference from $\tau=0$). The phase imparted by the
material for our model is again, $\phi(\omega) = \epsilon
\left(\omega - \omega_0\right)^2$.  Thus, we can write the
field after the dispersive sample as
\begin{equation}
E_2(\omega,\tau) = \left[E(\omega;A)-E(\omega;-A)\right] e^{i
\epsilon \left(\omega - \omega_0\right)^2 },
\end{equation}
where the minus sign reflects the $\pi$ phase shift acquired at the input
beam splitter.

It has been shown that pairs of strongly oppositely-chirped
laser pulses produce very narrow band sum-frequency generation
\cite{Raoult1998}~\hspace{-1mm}.  In the following calculation,
we make the assumption that the laser pulses have large chirp,
such that the pulse durations is much longer than the coherence
time (or transform limited pulse duration) and that the
difference in the delay between the two paths in the
interferometer are small compared with the chirped pulse
duration. In this limit, the SFG from the cross-correlation
(Fig.~2b) will have much narrower bandwidth than the
autocorrelation (Fig.~2c) and thus can be selected using
spectral filtering allowing us to ignore the contribution from
the autocorrelation.  In this case, the CPI signal will be
given by

\begin{equation}
E_3(\omega,\tau)\approx\int d\omega' \left[E(\omega';-A)
E(\omega-\omega';A) - E(\omega';A)E(\omega-\omega';-A)\right] e^{i
\tau \omega' + i \epsilon \left(\omega - \omega' - \omega_0\right)^2
}
\end{equation}

The total SFG power detected as a function of the time delay $\tau$
is given by,
\begin{eqnarray}
\label{CPIsignal}
I_{SFG}(\tau) & \propto & \int d\omega \left|E_3(\omega,\tau)\right|^2 \nonumber\\
& \propto & \Lambda_+ e^{-\frac{\tau^2}{2 \tau_+^2}}+ \Lambda_-
e^{-\frac{\tau^2}{2 \tau_-^2}}- \Lambda_c \cos\left[\zeta - \alpha
\tau^2 \right] e^{-\frac{\tau^2}{2\tau^2_{cpi}}},
\end{eqnarray}
with the following definitions,
\begin{equation}
\begin{array}{ccccccc}
\Lambda_\pm & = & \frac{\left(2 \pi^5 \right)^\frac{1}{2}\sigma}
{\left[1+2 \left(2 A^2\pm 2 A \epsilon + \epsilon^2\right) \sigma^4
\right]^\frac{1}{2}}, & \qquad & \Lambda_c & = & \frac{\left(2 \pi^5
\right)^\frac{1}{2}\sigma} {\left\{16 A^2 \epsilon^4 \sigma^{12}+
\left[1+2 (2 A^2+\epsilon^2)\sigma^4\right]^2\right\}^\frac{1}{4}},\\
\zeta & = & \frac{1}{2} \arctan\left[\frac{4 A \epsilon^2 \sigma^6}
{1+2\left(2 A^2+\epsilon^2\right)\sigma^4}\right],
& \qquad &
\alpha & = & \frac{2 A \epsilon^2 \sigma^8}{4 A^2\sigma^4 +
\left(1+2\epsilon^2\sigma^4\right)^2},\\
\tau_\pm & = & \left[\frac{1}{\sigma^2}+2\left( 2 A^2 \pm 2 A
\epsilon + \epsilon^2 \right)\sigma^2\right]^\frac{1}{2} &
 & \tau_{cpi} & = & \frac{1}{\sigma}\left[ \frac{4
A^2\sigma^4 + \left(1+2\epsilon^2\sigma^4\right)^2}{ 1+2\left(2 A^2
+ \epsilon^2\right)\sigma^4 } \right]^\frac{1}{2}.
\end{array}
\end{equation}

This are rather complicated expressions, so it is worth
examining the terms in a little detail.  The first two terms in
Eq.~\ref{CPIsignal} are nearly identical when $A\gg\epsilon$
and describe a broad Gaussian signal with the time duration of
the chirped pulses. The last term describes the interference
dip. If the chirp is large compared to the dispersion, the
cosine will be $\approx 1$ where the CPI dip occurs. Outside of
this limit the cosine term will start to modulate the
interference term, and the phase offset $\zeta$ will start to
reduce the visibility of the CPI dip.

The RMS width of the CPI dip is given by the expression for
$\tau_{cpi}$ which can be written
\begin{eqnarray}
\tau_{cpi} & = & \frac{1}{\sigma}\sqrt{ 1+ \epsilon^2 \sigma^4
\left(\frac{2+4\epsilon^2
\sigma^4}{1+4A^2\sigma^4+2\epsilon^2\sigma^4}\right)},
\end{eqnarray}
which has a similar form to Eq.~\ref{approxHOMwidth}. When the
chirp is larger than the dispersion, we can simplify this
expression to,
\begin{eqnarray}
\tau_{cpi} & = & \frac{1}{\sigma}\sqrt{ 1+ \epsilon^2 \sigma^4
\left(\frac{2+4\epsilon^2 \sigma^4}{1+4A^2\sigma^4}\right)}.
\end{eqnarray}
The dispersion in CPI is reduced by a small factor,
$\frac{2+4\epsilon^2 \sigma^4}{1+4A^2\sigma^4}$.  This ratio is
approximately the ratio
$\frac{\tau_\epsilon^2}{\tau_{chirp}^2}$, where $\tau_\epsilon$
is the time duration of a transform-limited pulse subjected to
the dispersive phase, $\phi(\omega)$, and $\tau_{chirp}$ is the
time duration of the chirped pulse.  Here, in strong analogy
with the quantum case, we can view the dispersion as having
been reduced from $\epsilon$ to
$\epsilon\frac{\tau_\epsilon}{\tau_{chirp}}$.  Thus as the
chirp pulse duration goes to infinity, the dispersion
cancellation becomes perfect.  With grating-based stretchers
and compressors, femtosecond pulses can easily be stretched
\cite{Pessot1987} by a factor of 1000, thus reducing the
sensitivity to the typically dominant second-order dispersion
by two to three orders of magnitude.

\section{CONCLUSIONS}

We have presented models describing the entangled photons in
Hong-Ou-Mandel interferometry and the laser pulses in
chirped-pulse interferometry which allow for varying degrees of
correlation.  In the case of perfect correlations, both HOM and
CPI exhibit complete insensitivity to second-order dispersion
\cite{evenorderfootnote}~\hspace{-1mm}.  When imperfect
correlations are considered, neither interferometer completely
cancels second-order dispersion, but the effective second-order
dispersion can be reduced significantly.  In the HOM
interferometer, the second-order dispersion is reduced by
roughly the ratio of the down-converted photon coherence time
to the pump coherence time.  In CPI, the dispersion is reduced
by the ratio of the time duration of a transform-limited pulse
subjected to the dispersive phase to the chirped pulse
duration.  It is straightforward in practice to make both of
these ratios small, although it is certainly easier to make the
quantum ratio very small ($<10^{-6}$).  For thick enough
materials or extremely large bandwidth, odd-order dispersion
terms, which are not cancelled, will dominate the signal after
which further reduction in the sensitivity to second-order
dispersion will not help signal quality.

Our results constitute simple rules of thumb for estimating the
maximum tolerable level of dispersion in the HOM interferometer
and CPI for a given strength of correlations. These
considerations will be important in designing optical
technologies which exploit automatic dispersion cancellation
with either quantum or classical resources.

\acknowledgments

The authors thank Grant Salton, Kostadinka Bizheva, Donna
Strickland, and Gregor Weihs for valuable discussions.  This work
was supported by National Sciences and Engineering Research Council
of Canada, the Canadian Foundation for Innovation, Ontario Centres
of Excellence, and an Ontario Ministry of Research and Innovation
Early Researcher Award.  R.K. is partially funded by the Institute
for Quantum Computing; D.B. is funded by a Mike and Ophelia
Lazaridis Scholarship; J.L is partially funded by the Bell Family
Fund.


\begin{thebibliography}{99}

\bibitem{Michelson1887} Michelson, A.A. and Morley, E.W., ``On
    the Relative Motion of the Earth and the Luminiferous
    Ether'', American Journal of Science 34, 333-345 (1887).

\bibitem{Abbott2005} Abbott, B. et al., ``Upper limits on a
    stochastic background of gravitational waves,'' Phys. Rev.
    Lett. 95, 221101 (2005).

\bibitem{Steinberg1992a} Steinberg, A.M., Kwiat, P.G. and
    Chiao, R.Y., ``Dispersion cancellation in a measurement of the
    single-photon propagation velocity in glass,'' Phys. Rev.
    Lett. 68, 2421-2424 (1992).

\bibitem{Steinberg1992b} Steinberg, A.M., Kwiat, P.G. and Chiao, R.Y., ``Dispersion cancellation and high-resolution time
measurements in a fourth-order optical interferometer,'' Phys. Rev.
A 45, 6659-6665 (1992).

\bibitem{Franson1992} Franson, J.D., ``Nonlocal cancellation of dispersion,'' Phys. Rev. A 45, 3126-3132 (1992).

\bibitem{Hong1987} Hong, C.K., Ou, Z.Y., and Mandel, L., ``Measurement of subpicosecond time intervals
between two photons by interference,'' Phys. Rev. Lett. 59,
2044-2046 (1987).

\bibitem{Franson1989} Franson, J.D., ``Bell inequality for position and time,'' Phys. Rev. Lett. 62, 2205-2208 (1989).

\bibitem{Ou1990} Ou, Z.Y., Zou, X.Y., Wang, L.J., and Mandel, L., ``Observation of nonlocal interference in separated photon channels,'' Phys. Rev. Lett.
65, 321-324 (1990).

\bibitem{Pittman1995} Pittman, T.B., Shih, V.H., Strekalov, D.V., and
Sergienko, A.V., ``Optical imaging by means of two-photon quantum
entanglement,'' Phys. Rev. A 52, R3429-R3432 (1995).

\bibitem{Strekalov1995} Strekalov, D.V., Sergienko, A.V., Klyshko, D.N., and
Shih, Y.H., ``Observation of Two-Photon "Ghost" Interference and
Diffraction,'' Phys. Rev. Lett. 74, 3600-3603 (1995).

\bibitem{Lee2002} Lee, H., Kok, P., and Dowling, J.P., ``A quantum Rosetta stone for interferometry,''
J. Mod. Opt. 49, 2325-2338 (2002).

\bibitem{Giovannetti2004} Giovannetti, V., Lloyd, S. and Maccone, L., ``Quantum-Enhanced Measurements: Beating the Standard Quantum Limit,'' Science 306,
1330-1336 (2004).

\bibitem{Mitchell2004} M. W. Mitchell, J. S. Lundeen, and A. M. Steinberg, ``Super-resolving phase measurements with a multiphoton entangled state,'' Nature 429,
161-164 (2004).

\bibitem{Walther2004} P. Walther, J.-W. Pan, M. Aspelmeyer, R. Ursin, S. Gasparoni,
and A. Zeilinger, ``De Broglie wavelength of a non-local four-photon
state,'' Nature 429, 158-161 (2004).

\bibitem{Yurke1986} B. Yurke, ``Input states for enhancment of fermion interferometer sensitivity,'' Phys. Rev. Lett. 56, 1515-1517 (1986).

\bibitem{Rarity1990} J. G. Rarity, P. R. Tapster, E. Jakeman, T. Larchuk, R. A.
Campos, M. C. Teich, and B. E. A. Saleh, ``2-photon interference in
a Mach-Zehnder interferometer,'' Phys. Rev. Lett. 65, 1348-1351
(1990).

\bibitem{Bennink2002} R. S. Bennink, S. J. Bentley, and R.W. Boyd, `` `Two-photon' coincidence imaging with a classical source,'' Phys. Rev. Lett. 89,
113601 (2002).

\bibitem{Ferri2005} F. Ferri, D. Magatti, A. Gatti, M. Bache, E. Brambilla, and L.A. Lugiato,
``High-resolution ghost image and ghost diffraction experiments with thermal light,'' Phys. Rev. Lett. 94, 183602 (2005).

\bibitem{Resch2007} K.J. Resch, P. Puvanathasan, J.S. Lundeen, M.W. Mitchell, and K. Bizheva,
``Classical dispersion-cancellation interferometry,'' Opt. Express
15, 8797-8804 (2007).

\bibitem{Resch2007b} K.J. Resch, P. Puvanathasan, K. Bizheva, M. Mitchell, and J.S.
Lundeen, ``White-light interferometer automatically cancels
dispersion,'' Laser Focus World 43, 85-89 (2007).

\bibitem{Kaltenbaek2008} R. Kaltenbaek, J. Lavoie, D. N. Biggerstaff, and K. J.
Resch, ``Quantum-inspired interferometry with chirped laser pulses
,'' Nature Phys. 4, 864-868 (2008).

\bibitem{Lavoie2009} J. Lavoie, R. Kaltenbaek, and K. J. Resch, ``Quantum-optical coherence tomography with classical light,'' Opt. Express 17,
3818-3825 (2009).

\bibitem{Resch2007a} K.J. Resch, K.L. Pregnell, R. Prevedel, A. Gilchrist, G.J. Pryde,
J.L. O'Brien, and A.G. White, ``Time-reversal and super-resolving
phase measurements ,'' Phys. Rev. Lett. 98, 223601 (2007).

\bibitem{Kaltenbaek2009} R. Kaltenbaek, J. Lavoie, and K. J.
    Resch, ``Classical analogues of two-photon quantum interference,'' Phys. Rev. Lett. 102, 243601 (2009).

\bibitem{KFLee2002} K.F. Lee and J.E. Thomas, ``Experimental Simulation of Two-Particle
Quantum Entanglement using Classical Fields,'' Phys. Rev. Lett. 88,
097902 (2002).

\bibitem{KFLee2004} K.F. Lee and J.E. Thomas, ``Entanglement with classical fields,'' Phys. Rev. A 69, 052311
(2004).

\bibitem{KFLee2009} K.F. Lee, ``Observation of bipartite correlations using
coherent light for optical communication,''Optics Letters 34,
1099-1101 (2009).

\bibitem{Erkmen2006} B.I. Erkmen and J.H. Shapiro, ``Phase-conjugate optical coherence tomography,'' Phys. Rev. A 74, 041601 (2006).

\bibitem{Banaszek2007} K. Banaszek, A.S. Radunsky, and I.A. Walmsley, ``Blind dispersion compensation for optical coherence tomography,'' Opt. Commun. 269,
152-155 (2007).

\bibitem{Nasr2008} M.B. Nasr, S. Carrasco, B.E.A. Saleh, A.V.
Sergienko, M.C. Teich, J.P. Torres, L. Torner, D.S. Hum, M.M. Fejer,
``Ultrabroadband biphotons generated via chirped quasi-phase-matched
optical parametric down-conversion'' Phys. Rev. Lett. 100, 199903
(2008).

\bibitem{Abouraddy2002} A. F. Abouraddy, M. B. Nasr, B. E. A. Saleh, A. V. Sergienko, and
M. C. Teich, ``Quantum-optical coherence tomography with dispersion
cancellation,'' Phys. Rev. A 65, 053817 (2002).

\bibitem{Nasr2003} M.B. Nasr, B.E.A. Saleh, A.V. Sergienko, and M.C. Teich,
``Demonstration of dispersion-canceled quantum-optical coherence
tomography,'' Phys. Rev. Lett. 91, 083601 (2003).

\bibitem{Kwiat1993} P.G. Kwiat, A.M. Steinberg, and R.Y. Chiao,
``High-visibility interference in a Bell-inequality experiment for
energy and time,'' Phys. Rev. A 47, R2472-R2475 (1993).

\bibitem{BoydNLO} Boyd, R.W., [Nonlinear Optics Third Edition], Academic Press, New York
p.76 (2008).

\bibitem{Raoult1998} F. Raoult, A. C. L. Boscheron, D. Husson, C. Sauteret, A. Modena, V. Malka, F. Dorchies, and A. Migus,
``Efficient generation of narrow-bandwidth picosecond pulses by
frequency doubling of femtosecond chirped pulses,'' Opt. Lett. 23,
1117-1119 (1998).

\bibitem{Pessot1987} M. Pessot, P. Maine and G. Mourou,
``1000 times expansion compression of optical pulses for chirped
pulse amplification,'' Opt. Commun. 62, 419-421 (1987).

\bibitem{evenorderfootnote} In
    Ref.~\cite{Kaltenbaek2008}~\hspace{-1mm}, we showed that
    CPI is insensitive to all even orders of dispersion in the
    case of perfect correlation.

\end{thebibliography}
\end{document}